\documentclass[10pt]{article}
\usepackage{a4}
\begin {document}
\begin{title}
{
\hfill{\small {\bf MKPH-T-98-8}}\\
{\bf Invariant amplitudes for coherent electromagnetic pseudoscalar 
production from a spin-one target}
\footnote{Supported by the Deutsche 
Forschungsgemeinschaft (SFB 201)}}
\end{title}
\bigskip
\author{
Hartmuth Arenh\"ovel \\
Institut f\"ur Kernphysik\\ 
Johannes Gutenberg-Universit\"at\\
D-55099 Mainz, Germany}
\maketitle
\begin{abstract}
A set of 13 linearly independent invariant amplitudes for the 
electromagnetic production of a pseudoscalar particle from a spin-one 
particle is derived which respect Lorentz and gauge invariance. The 
$T$-matrix can be represented by a linear superposition of these amplitudes
with invariant functions as coefficients, which depend on the Mandelstam 
variables only. Their explicit form is determined by the underlying dynamics. 
Nine of these amplitudes are purely transverse and describe photoproduction. 
The remaining four appear in electroproduction in addition, describing 
charge and longitudinal current contributions. Furthermore, a reduction
of these amplitudes to operators acting in non-relativistic spin space is
given. 
\end{abstract}
\section{Introduction}
In recent years, coherent pseudoscalar meson production in electromagnetic 
reactions on nuclei has become a very active field of research in 
medium energy physics. 
A particularly interesting role plays the deuteron, since it is the simplest 
nucleus on whose structure we have abundant information and a reliable 
theoretical understanding. Morover, it constitutes the simplest and 
cleanest neutron target, and thus allows one to study neutron 
properties, provided one has control on binding effects in the most general 
sense. In fact, this is one main motivation for studying coherent pion and 
eta photo- and electroproduction on the deuteron, both in experiment and 
theory (see, for example \cite{BlB94,WiA95,KaT97,BrA97} and references 
therein). 
Most theoretical approaches thus start from the one-body contributions to the 
reaction matrix which, therefore, is given by specific contributions 
in terms of the nucleon variables. However, it might be useful to investigate
the general structure of the reaction amplitude without any specific input
from the internal degrees of freedom of the target, using only the general 
principles of basic conservation laws, like Lorentz covariance, gauge 
invariance etc. In other words, we want to determine the most general framework
for the formal description of the reaction matrix into which any specific 
reaction model has to fit. This is in analogy to the CGLN-amplitudes for pion 
photo- and electroproduction on a nucleon \cite{CGLN,Don72}. 

To this end, we first construct in Sect.\ 2 the most general form of the 
transition amplitude in terms of basic amplitudes which respect Lorentz 
covariance and parity conservation. Then, in Sect.\ 3 we determine from 
these basic amplitudes a set of 13 linearly independent amplitudes which 
in addition respect gauge invariance. Helicity amplitudes are derived in 
Sect.\ 4. They serve as a very convenient basis for the construction of 
observables. Finally, in the last section we 
relate these amplitudes to a corresponding basic set of 13 nonrelativistic 
operators acting in spin-one spinor space. 

\section{Formal developments}
The basic reaction, coherent pseudoscalar 
production by photoabsorption or electron scattering from a spin-one target, 
is of the form 
\begin{eqnarray}
\gamma^{(*)}(k) + a(p) \rightarrow a(p') + b(q)\,, 
\end{eqnarray}
where the real or virtual photon has momentum $k$, $a$ is a spin-one 
particle of mass $M$ with initial and final momenta $p$ and $p'$, 
respectively, and $b$ a pseudoscalar meson of mass $m$ with momentum $q$. 
In the c.m.\ frame of photon and target particle, the differential cross 
section for electromagnetic production with real or virtual photons is in 
general given by
\begin{eqnarray}
d\sigma^{c.m.}_{fi}=c_\gamma\,tr\,({\cal M}_{fi}^\dagger {\cal M}_{fi}\,
\hat\rho^\gamma \hat\rho^{target})\,d\Omega^{c.m.}_q\,,
\end{eqnarray}
where $\hat\rho^\gamma$ and $\hat\rho^{target}$ denote the density matrices 
of the real or virtual photon and the target particle, respectively, and 
$c_\gamma$ contains corresponding kinematic factors whose specific forms 
are not needed for the present study. The trace is to be taken with respect 
to all polarization quantum numbers. 

${\cal M}_{fi}$ denotes the invariant matrix element which has to be 
linear in the polarization vectors of the participating particles. 
Therefore, it will have the general structure  
\begin{eqnarray}
{\cal M}_{fi}=U^{\mu\dagger}(\vec p^{\,\prime},m')\, O_{\mu\nu\rho}(p',q,p,k) 
\,\epsilon^\nu(k,\lambda)\,U^\rho(\vec p,m)\,,
\end{eqnarray}
where the initial and final polarization Lorentz vectors of the spin-one 
particle with spin projections $m$ and $m'$ are denoted by $U^\rho(\vec p,m)$ 
and $U^\mu(\vec p^{\,\prime},m')$, respectively. The explicit form of the 
polarization vector $U(\vec p,m)$ of a massive spin-one particle is given by
\begin{eqnarray}
U^\mu(\vec p,m)=\frac{1}{M}\,(\vec p \cdot \vec e_m, M\vec e_m+ 
\frac{\vec p \cdot \vec e_m}{E_p+M}\,\vec p \,)\,,\label{Uspinor}
\end{eqnarray}
where $E_p=\sqrt{\vec p^{\,2}+M^2}$, and the rest frame polarization 
three-vector is denoted by $\vec e_m$. $U(\vec p,m)$ obeys the 
transversality condition
\begin{eqnarray}
p_\rho \,U^\rho(\vec p,m)=0\,.\label{tansvcond}
\end{eqnarray}
The photon polarization vectors are denoted by $\epsilon^\nu(k,\lambda)$. 
For real photons one has only two transverse polarization states with helicity 
$\lambda=\pm 1$ which have only spatial components 
\begin{eqnarray}
\epsilon^\nu(k,\lambda=\pm 1)=(0,\vec \epsilon_{\pm 1})\, ,\label{photpolt}
\end{eqnarray} 
with $\vec \epsilon_{\pm 1} \cdot \vec k=0$ and $|\vec \epsilon_{\pm 1}|=1$, 
whereas for electroproduction in addition a third polarization state 
($\lambda=0$) appears having both, longitudinal and timelike components 
according to
\begin{eqnarray}
\epsilon^\nu(k,\lambda=0)=
\frac{|\vec k|}{\sqrt{K^2}}\,(1,\frac{k_0}{|\vec k|^2}
\vec k)\,,\label{photpoll}
\end{eqnarray}
where we have introduced for convenience $K^2=-k^2_\mu>0$. 
In both cases, one has the transversality condition
\begin{eqnarray}
k_\nu\epsilon^\nu(k,\lambda) = 0\,.\label{transk}
\end{eqnarray}
Furthermore, assuming parity conservation, $O_{\mu\nu\rho}$ has to be a 
third rank Lorentz pseudotensor since a pseudoscalar is produced. 
Folded between the target polarization Lorentz vectors, it describes the axial
current matrix elements for the coherent pseudoscalar production
\begin{eqnarray}
j^\nu_{fi}(p',q,p,k)=U_{\mu}^{\dagger}(\vec p^{\,\prime},m')\, 
O^{\mu\nu\rho}(p',q,p,k)\,U_\rho(\vec p,m)\,.
\end{eqnarray}
As constraint from gauge invariance one has the condition of 
current conservation
\begin{eqnarray}
k_\nu \,j^\nu_{fi}(p',q,p,k)=U_{\mu}^{\dagger}(\vec p^{\,\prime},m') 
\,O^{\mu\nu\rho}(p',q,p,k)\, k_\nu \,U_\rho(\vec p,m)=0\,.\label{gauge}
\end{eqnarray}

The operator $O_{\mu\nu\rho}$ has 
to be constructed from the momenta of the participating particles, 
from the metric tensor 
$g^{\mu\nu}$ and the four-dimensional completely antisymmetric Levy-Civita
tensor $\varepsilon^{\mu\nu\rho\sigma}$. The presence of the latter is 
necessary in order to guarantee the pseudotensor property.
We will now construct a set of basic covariant amplitudes $\{\Omega_\alpha\}$ 
which serves as a basis for the representation of the invariant matrix 
element as a linear superposition of the basic amplitudes, i.e., 
\begin{eqnarray}
{\cal M}_{fi}=\sum_\alpha F_\alpha(s,t,u)\,\Omega_\alpha\,, \label{mficov}
\end{eqnarray}
where $F_\alpha(s,t,u)$ denote invariant functions which depend solely on the 
Mandelstam variables $s$, $t$ and $u$, defined as usual by
\begin{eqnarray}
s = (k+p)^2\,,\quad 
t = (k-q)^2\,,\quad
u = (k-p'\,)^2\,.
\end{eqnarray}
Because of the condition 
\begin{eqnarray}
s+t+u= 2M^2-K^2 +m^2\,,
\end{eqnarray}
only two of them are independent, 
for example, $s$ and $t$. The specific form of the invariant functions 
$F_\alpha(s,t,u)$ will 
depend on the detailed dynamical properties of the target and the produced 
meson. Because of momentum conservation
\begin{eqnarray}
k+p=p'+q\,,
\end{eqnarray}
only three of the particle momenta can be considered as independent. We 
will choose in the following $k,\,p$ and $q$. 

Counting the number of polarization states for the initial and final 
particles,
one finds  for electroproduction a total number of $T$-matrix elements of 
$3\times 3 \times 3=27$ and for photoproduction 18, which reduces to 13, 
respective 9 independent $T$-matrix elements, if parity conservation can 
be invoked (see for example \cite {WiA95}). Consequently, one has to find 
13 respective 9 linearly independent invariant amplitudes. By this we mean 
that a set $\{\Omega_\alpha\}$ is called linearly independent if one 
cannot find a set of scalar functions $\{f_\alpha(s,t,u)\}$ such that 
the equation
\begin{eqnarray}
\sum_\alpha f_\alpha(s,t,u)\,\Omega_\alpha\equiv 0 
\end{eqnarray}
holds identically. 

First we will select all possible amplitudes for $O^{\mu\nu\rho}$ respecting 
the requirements of Lorentz covariance and parity conservation. 
In view of the pseudotensor character, each amplitude has to be linear in the 
Levy-Civita tensor, because higher order odd powers of 
$\varepsilon^{\mu\nu\rho\sigma}$ can be reduced to linear expressions
in $\varepsilon^{\mu\nu\rho\sigma}$ (see Appendix A). Any two amplitudes,
which only differ by scalar products between the vectors $k,\,p$ and $q$, 
cannot be considered as independent, because the scalar products can be 
expressed in terms of the Mandelstam variables according to
\begin{eqnarray}
k\cdot p &=&\frac{1}{2}\,(s-M^2+K^2)\,,\\
k\cdot q &=&\frac{1}{2}\,(m^2-K^2-t)\,,\\
p\cdot q &=&\frac{1}{2}\,(s+t-M^2+K^2)\,.
\end{eqnarray}
For the further discussion it is useful to 
introduce for convenience a covariant pseudoscalar by contraction of 
$\varepsilon^{\mu\nu\rho\sigma}$ with four Lorentz vectors $a,\,b,\,c$ 
and $d$ 
\begin{eqnarray}
S(a,b,c,d)=\varepsilon^{\mu\nu\rho\sigma}a_\mu b_\nu c_\rho d_\sigma\,.
\end{eqnarray}
Any amplitude thus has to contain such a pseudoscalar, and one can distinguish
the possible types according to whether one, two or three of the 
polarization vectors are contained in $\varepsilon^{\mu\nu\rho\sigma}$. 
In other words, the possible candidates for $O^{\mu\nu\rho}$ are
\begin{eqnarray}
&\varepsilon^{\mu\nu\rho\sigma}x_\sigma,\,&\nonumber\\
\varepsilon^{\mu\nu\sigma\tau}x_\rho y_\sigma z_\tau,\,&
\varepsilon^{\mu\rho\sigma\tau}x_\nu y_\sigma z_\tau,\,&
\varepsilon^{\nu\rho\sigma\tau}x_\mu y_\sigma z_\tau,\,\nonumber\\
\varepsilon^{\mu\lambda\sigma\tau}g_{\nu\rho} z_\lambda u_\sigma v_\tau,\,&
\varepsilon^{\nu\lambda\sigma\tau}g_{\mu\rho} z_\lambda u_\sigma v_\tau,\,&
\varepsilon^{\rho\lambda\sigma\tau}g_{\mu\nu} z_\lambda u_\sigma v_\tau\,,
\nonumber\\
\varepsilon^{\mu\lambda\sigma\tau}x_\nu y_\rho z_\lambda u_\sigma v_\tau,\,&
\varepsilon^{\nu\lambda\sigma\tau}x_\mu y_\rho z_\lambda u_\sigma v_\tau,\,&
\varepsilon^{\rho\lambda\sigma\tau}y_\mu y_\nu z_\lambda u_\sigma v_\tau\,,
\nonumber
\end{eqnarray}
where $x,y,z,u,v \in \{k,p,q\}$. Since only three independent 
kinematic vectors are 
available, no pseudoscalar of the type $S(x,y,z,u)$ will appear. Therefore, 
one finds as basic types of amplitudes the ones listed in 
Table \ref{tab0}, where we use as a shorthand $U'$ for 
$U^\dagger(\vec p^{\,\prime},m')$ and $U$ for $U(\vec p,m)$, 
Note, that
in all following expressions the polarization vector $U'$ of the final 
target state always has to stand on the left hand side with respect to $U$.
Taking into account the fact, that $S(a,b,c,d)$
vanishes if two arguments are equal, one finds three different amplitudes for 
each type $\Omega_a$ and $\Omega_e$, while for each of the others one finds 
nine. Thus the total number of possible amplitudes is 60, too many according
to what has been said above about the number of independent amplitudes. 

\section{Construction of independent invariant amplitudes}
The main task now is to determine from the set of amplitudes in 
Table \ref{tab0} the independent
ones. A first reduction is achieved by taking into account the transversality 
conditions (\ref{tansvcond}) and (\ref{transk}) from which follows
\begin{eqnarray}
  \Omega_b(x, y,p)&=& 0\,,\\ 
  \Omega_d(x, y,k)&=& 0\,,\\ 
  \Omega_f(x,p)&=&0\,,\\ 
  \Omega_f(k,y)&=&0\,,\label{transfk}\\ 
  \Omega_g(x,k)&=&0\,,\label{transgk}\\ 
  \Omega_h(x,p)&=&0\,,
\end{eqnarray}
and, because of $U'\cdot p=U'\cdot q - U'\cdot k$, one furthermore has 
\begin{eqnarray}
  \Omega_c(x, y,p)&=&\Omega_c(x, y,q)-\Omega_c(x, y,k)\,,\\ 
  \Omega_{g/h}(p,y)&=&\Omega_{g/h}(q,y)-\Omega_{g/h}(k,y)\,.
\end{eqnarray}
This leaves 36 amplitudes from which we have to find the 
independent ones which in addition respect the gauge condition of 
(\ref{gauge}). 

A further reduction follows from
two linear relations for the Levy-Civita tensor (see the Appendix A). As 
is shown in detail in Appendix B, by
means of these relations one can eliminate the amplitudes 
$\Omega_a(k)$ and $\Omega_{b/c}(p,q,q)$ (see the relations (\ref{relqc}), 
(\ref{relqb}) and (\ref{omegae2}) of Appendix B). Furthermore, one can 
eliminate $\Omega_{b/c}(p,q,k)$, $\Omega_d$, $\Omega_f$, and $\Omega_g$  
according to the relations 
\begin{eqnarray}
\Omega_b(p,q,k)&=&-\Omega_b(k,p,q)
-\Omega_e(U',\epsilon,U)+\Omega_e(\epsilon,U',U)\,,\\
\Omega_c(p,q,k)&=&-\Omega_c(k,p,q)-\Omega_c(q,k,p)
-\Omega_e(U,U',\epsilon)+\Omega_e(\epsilon,U',U)\,,\\
\Omega_d(x,y,z)&=&\Omega_a(x)\,y\cdot z-\Omega_a(y)\,x\cdot z 
+\Omega_b(x,y,z)+\Omega_c(x,y,z)\,,\\
\Omega_f(x,y)&=&\Omega_h(x,y)+Q_b(k,p,q;x,y)\,,\label{relf}\\
\Omega_g(x,y)&=&\Omega_h(x,y)-Q_c(k,p,q;y,x)\,,\label{relg}
\end{eqnarray}
where $Q_{b/c}$ is defined in (\ref{defqbc}) of Appendix B.

Of the remaining 17 linearly independent amplitudes, 13 ones are gauge 
invariant. They are listed in Table \ref{tab1}. It means that
these amplitudes vanish under the replacement $\epsilon \rightarrow k$. 
The other four non-gauge invariant amplitudes are listed in 
Table \ref{tab1a}. They cannot be linearly combined 
to form gauge invariant amplitudes and, therefore, cannot contribute to 
${\cal M}_{fi}$. In view of the above mentioned 13 independent $T$-matrix 
elements, it is not surprising that we find exactly 13 linearly 
independent invariant amplitudes
which respect gauge invariance and which could serve as a complete 
basis for ${\cal M}_{fi}$. However, not all of them have the most 
convenient form. While the first nine amplitudes of Table \ref{tab1} 
are purely transverse in the c.m.\ frame of photon and target particle, 
and, therefore, suffice and are well suited for describing 
photoproduction, the remaining four, which in addition are needed in 
electroproduction, have besides charge and longitudinal 
current components also transverse current pieces. For this reason it is 
more advantageous to replace the last four amplitudes of Table \ref{tab1} 
by equivalent amplitudes which are purely longitudinal in the c.m.\ frame. 
This is easily achieved by using the relations in (\ref{relf}) and 
(\ref{relg}) in conjunction with (\ref{transfk}) and (\ref{transgk}), yielding
\begin{eqnarray}
k\cdot p\, \Omega_f(k,k) - k_\mu^2\,\Omega_f(p,k)
&=&(k_\mu^2+k\cdot p\, \Omega_h(k,k)
- k_\mu^2\,\Omega_h(q,k)\label{equiv1}\nonumber\\
&& -[k,p;k,q]\,\Omega_b(k,p,k) + [k,p;k,p]\,\Omega_b(k,q,k)\,,\\
k\cdot p\, \Omega_f(k,q) - k_\mu^2\,\Omega_f(p,q)
&=&(k_\mu^2+k\cdot p\, \Omega_h(k,q)
- k_\mu^2\,\Omega_h(q,q) \nonumber\\
&& -[k,p;k,q]\,\Omega_b(k,p,q) + [k,p;k,p]\,\Omega_b(k,q,q)\nonumber\\
&& +2\,[k,p;k,q]\,\Omega_e(\epsilon,U',U)\,,\\
k\cdot p\, \Omega_g(k,k) - k_\mu^2\,\Omega_g(k,p)
&=&(k_\mu^2+k\cdot p\, \Omega_h(k,k)
- k_\mu^2\,\Omega_h(k,q) \nonumber\\
&& +[k,p;k,q]\,\Omega_c(k,p,k) - [k,p;k,p]\,\Omega_c(k,q,k)\,,\\
k\cdot p\, \Omega_g(q,k) - k_\mu^2\,\Omega_g(q,p)
&=&(k_\mu^2+k\cdot p\, \Omega_h(q,k)
- k_\mu^2\,\Omega_h(q,q) \nonumber\\
&& +[k,p;k,q]\,\Omega_c(k,p,q) - [k,p;k,p]\,\Omega_c(k,q,q)\,,\label{equiv4}
\end{eqnarray}
where the symbol $[a,b;c,d]$ is defined in (\ref{defklammer}).
Therefore, we can replace the last four amplitudes of Table \ref{tab1} by 
the equivalent ones as given on the left hand side of (\ref{equiv1}) through 
(\ref{equiv4}) and thus use the operators in Table \ref{tab1b} for the 
longitudinal contributions in electroproduction.

\section{Helicity amplitudes}

The most convenient framework for the description of observables like 
differential cross section, beam and target asymmetries and recoil 
polarization, is given by the helicity amplitudes in the c.m.\ frame of 
the reaction. For the coordinate system we choose the $z$-axis along the photon
momentum $\vec k$, the $y$-axis perpendicular to the plane built by photon 
and meson momentum in the direction $\vec k\times\vec q$ and finally the 
$x$-axis along $\vec e_y\times\vec e_z$. The corresponding spherical basis 
will be denoted by $\vec e_\lambda$ with $\lambda=0,\,\pm1$. The helicity 
states of the photon polarization are already given in (\ref{photpolt}) 
and (\ref{photpoll}). 
Similarly, the helicity states of the initial and final deuteron 
polarization are given by
\begin{eqnarray}
U^\mu(\vec p,\lambda)&=&\frac{(-)^\lambda}{M}\,\Big(k\delta_{\lambda 0},
                        -E(\lambda)\,\vec e_{-\lambda} \Big)\,,\\
U^\mu(\vec p^{\,\prime},\lambda')&=&\frac{(-)^{\lambda'}}{M}\,
    \Big(q\delta_{\lambda' 0},
    -E'(\lambda')\,\sum_{\bar \lambda}\vec e_{\bar\lambda}
     d^1_{\bar\lambda,-\lambda}(\theta) \Big)\,,
\end{eqnarray}
where $\theta$ denotes the angle between $\vec k$ and $\vec q$. Use has been made of the c.m.\ frame relations
\begin{eqnarray}
p^\mu=(E_k, -k \vec e_0),\quad p^{\prime \mu}= \Big(E_q, -q 
\sum_{\bar \lambda}\vec e_{\bar \lambda}d^1_{\bar \lambda 0}(\theta)\Big)\,.
\end{eqnarray}
In the c.m.\ frame the absolute values of the momenta $k=|\vec k\,|$ and 
$q=|\vec q\,|$ are given by the well-known expressions
\begin{eqnarray}
k^2&=&\frac{1}{4s}\Big((\sqrt{s}+M)^2+K^2\Big)
\Big((\sqrt{s}-M)^2+K^2\Big)\nonumber\\
&=&\frac{1}{4s}\Big((s-M^2+K^2\Big)^2+4K^2M^2\Big)\,,\\
q^2&=&\frac{1}{4s}\Big((\sqrt{s}+M)^2-m^2\Big)\Big((\sqrt{s}-M)^2-m^2\Big)
\nonumber\\
&=&\frac{1}{4s}\Big((s-M^2-m^2\Big)^2-4m^2M^2\Big)\,.
\end{eqnarray}
The rotation matrices $d^j_{m'm}$ are in the convention of \cite{Ros57}. 
Furthermore we have introduced in the above expressions the shorthand notation 
\begin{eqnarray}
E(\lambda)&=& M\delta_{|\lambda|1} +E_k\delta_{\lambda 0}\,,\\
E'(\lambda')&=& M\delta_{|\lambda'|1} +E_{q}\delta_{\lambda' 0}\,.
\end{eqnarray}

Explicit evaluation of the various terms in the expressions for the invariant 
transverse and longitudinal amplitudes listed in Tables \ref{tab1} and 
\ref{tab1b} yields
\begin{eqnarray}
U'_{\lambda'}\cdot U_\lambda &=&\frac{1}{M^2}\,
   \Big(kq\delta_{\lambda'0}\delta_{\lambda 0}-E'(\lambda')E(\lambda)
   d^1_{\lambda \lambda'}(\theta)\Big)\,,\\
U_\lambda\cdot k  &=&\frac{k}{M}\sqrt{s}\,\delta_{\lambda 0}\,,\\
U_\lambda\cdot q  &=&\frac{1}{M}\Big(kq_0\delta_{\lambda 0}+qE(\lambda)
   d^1_{\lambda 0}(\theta)\Big)\,,\\
U'_{\lambda'}\cdot k  &=&\frac{1}{M}\Big(qk_0\delta_{\lambda' 0}+kE'(\lambda')
   d^1_{0 \lambda'}(\theta)\Big)\,,\\
U'_{\lambda'}\cdot q  &=&\frac{1}{M}\,q\sqrt{s}\,\delta_{\lambda' 0}\,,\\
S(\epsilon_{\lambda_\gamma},k,p,q) &=& -i\lambda_\gamma kq\sqrt{s}\,
   d^1_{\lambda_\gamma 0}(\theta)\,,\\
S(U_{\lambda},k,p,q) &=& -i\lambda kq\sqrt{s}\,
   d^1_{\lambda 0}(\theta)\,,\\
S(U'_{\lambda'},k,p,q) &=& i\lambda' kq\sqrt{s}
   \,d^1_{\lambda' 0}(\theta)\,,\\
S(\epsilon_{\lambda_\gamma},U_{\lambda},k,p) &=&\frac{i}{M}\,\lambda E(\lambda)
   k\sqrt{s} \,\delta_{\lambda_\gamma\lambda}\,,\\
S(U'_{\lambda'},\epsilon_{\lambda_\gamma},k,p) &=& \frac{i}{M}
   \lambda_\gamma E'(\lambda') k\sqrt{s} 
   \,d^1_{-\lambda' \lambda_\gamma }(\theta)\,,\\
k\cdot p\,\,k\cdot\epsilon_{\lambda_\gamma} 
-k\cdot k\,\,p\cdot\epsilon_{\lambda_\gamma} &=&
   \sqrt{K^2}\,k\sqrt{s}\,\delta_{\lambda_\gamma 0}\,.
\end{eqnarray}
Note that on the rhs we have denoted the absolute values of the photon 
and meson three momenta by $k$ respective $q$. 
From these expressions it is straightforward to construct the helicity 
representation of the invariant amplitudes $\Omega_\alpha$. They are 
listed in Table \ref{tabhelicity}. With the help of these expressions 
the helicity representation of 
the $T$-matrix is obtained easily in terms of the invariant functions 
$F_\alpha(s,t,u)$ according to (\ref{mficov}). Note that the helicity 
amplitudes and thus the full $T$-matrix obey the symmetry relation 
which reflects parity conservation 
\begin{eqnarray}
\Omega_{\alpha, -\lambda' -\lambda_\gamma -\lambda}=
(-)^{1+\lambda'+ \lambda_\gamma +\lambda}
\Omega_{\alpha, \lambda' \lambda_\gamma \lambda}\,,
\end{eqnarray}
as is evident from the explicit expressions in Table \ref{tabhelicity}.

\section{Representation with respect to nonrelativistic spin-one spinors}

For the description of ${\cal M}_{fi}$ in the c.m.\ frame with respect to 
the nonrelativistic spin-one spinor space, we will first determine a basis 
of linearly independent longitudinal and transverse operators in terms of 
the only available momenta $\vec k$ and $\vec q$ 
and the spin-one operators $S^{[S]}$ $(S=0,1,2)$ of rank zero, one and 
two, using the notation of Fano and Racah \cite{FaR59}. Explicitly, 
$S^{[0]}=I_3$ is the three-dimensional unit matrix, 
$S^{[1]}$ the usual spin operator and $S^{[2]}$ is defined by 
\begin{eqnarray}
S^{[2]}=[S^{[1]}\times S^{[1]}]^{[2]}\,.
\end{eqnarray}
Here, the symbol $[\vec a\times \vec b\,]^{[2]}$ stands for the coupling 
of two vectors $\vec a$ and $\vec b$ to a second-rank tensor. It is defined by
\begin{eqnarray}
[\vec a\times \vec b\,]^{[2]}_{kl}:=\frac{1}{2}(a_k b_l +a_l b_k)
-\frac{1}{3}\, \delta_{kl}\, \vec a \cdot \vec b\,.
\end{eqnarray}
The coupling of a vector $\vec a$ with itself will be denoted by the shorthand 
$a^{[2]}:=[\vec a\times \vec a\,]^{[2]}$. 
The longitudinal and transverse operators must have the form 
\begin{eqnarray}
{\cal O}_L=x\,,\qquad 
{\cal O}_T=\vec \epsilon\cdot (\hat k \times \vec y)\,,
\end{eqnarray}
where $x$ is a pseudoscalar and $\vec y$ a polar vector to be constructed
from $\hat k$, $\hat q$ and $S^{[S]}$, denoting the unit vector of $\vec v$
by $\hat v =\vec v/|\vec v|$. We have four pseudoscalars available
\begin{eqnarray}
\hat k \cdot \vec S,\,\,\hat q \cdot \vec S,\,\,
[(\hat k\times \hat q\,)\times \hat k]^{[2]}\cdot S^{[2]},\,\,\mbox{and } 
[(\hat k\times \hat q\,)\times \hat q\,]^{[2]}\cdot S^{[2]}\,. 
\end{eqnarray}
 
For the vector $\vec y$ of the transverse operators one has two types: first 
$\hat q$ times a scalar for which the following types are available 
\begin{eqnarray}
1,\,\,(\hat k \times \hat q\,) \cdot\vec S,\,\,
[\hat k\times \hat q\,]^{[2]}\cdot S^{[2]},\,\,
\hat k^{[2]}\cdot S^{[2]},\,\,\mbox{and } \hat q^{[2]}\cdot S^{[2]}\,, 
\end{eqnarray}
and second 
\begin{eqnarray}
\hat u \times \vec S\,\,\mbox{and }[\hat u \times S^{[2]}]^{[1]},\,\,
\mbox{with }\hat u \in \{\hat k, \hat q\}\,,
\end{eqnarray}
where the coupling of a vector with a second-rank tensor to form a vector 
is given by
\begin{eqnarray}
[\vec u \times S^{[2]}]^{[1]}_k=u_lS^{[2]}_{lk}\,.  
\end{eqnarray}

The resulting thirteen basic operators are listed in Table 
\ref{tabkqls}. One should note, that the first four transverse and the 
first two longitudinal operators correspond to the CGLN-operators for a 
spin one-half target. The additional operators appear due to the presence 
of the second rank spin tensor $S^{[2]}$.
All other operators, which can be constructed 
from $\vec k$ and $\vec q$ 
and the spin-one operators $S^{[S]}$, can be reduced to linear 
combinations of the above basic operators with scalar functions of 
$\vec k^{\,2}$, $\vec q^{\,2}$, and $\vec k \cdot \vec q$ as coefficients, 
which in turn may be expressed as functions of the Mandelstam variables. 

In complete analogy to the CGLN-amplitudes in e.m.\ pion production, one can 
expand the general $T$-matrix ${\cal M}_{fi}$ in terms of the basic 
operators of Table \ref{tabkqls}, denoting the longitudinal and transverse 
parts by a proper superscript
\begin{eqnarray}
{\cal M}_{fi}^{L/T}=\chi_{m'}^{[1]\,\dagger}
\Big(\sum_{\beta} G_{\beta}^{L/T}(s,t,u)\,
{\cal O}_{L/T,\beta}\,\Big) 
\chi_{m}^{[1]}\,,
\label{gnonrel}
\end{eqnarray}
where again $G_{\beta}^{L/T}(s,t,u)$ are invariant functions, and 
$\beta=1,\dots,4$ for the longitudinal and $\beta=1,\dots,9$ for the 
transverse matrix elements. The nonrelativistic spin-one spinors are 
denoted by $\chi_{m}^{[1]}$. 

In order to relate this representation to the one in (\ref{mficov}),
we start from the general form of the polarization vector 
$U(\vec p,m)$ of a massive spin-one particle as given in (\ref{Uspinor}).
The rest frame polarization three-vector $\vec e_m$ is related to the 
nonrelativistic spin-one spinor $\chi_m^{[1]}$ with $z$-axis as quantization 
axis by
\begin{eqnarray}
\vec e_m=\vec P\chi_m^{[1]}\,.
\end{eqnarray}
The components of $\vec P$ are $(1\times 3)$-matrices 
\begin{eqnarray}
P_x=\frac{1}{\sqrt{2}}(-1,0,1),\quad
P_y=-\frac{i}{\sqrt{2}}(1,0,1),\quad
P_z=(0,1,0)\,,
\end{eqnarray}
where we note the orthogonality property $P_kP_l^\dagger=\delta_{kl}$.
It is straightforward to show that the bilinear products of the form 
$P_k^\dagger P_l$, which appear in the basic amplitudes, can be expressed
by the spin-one operators of rank zero, one and two according to 
\begin{eqnarray}
{P_k}^\dagger P_l=\frac{1}{3}\,\delta_{kl}I_3 +\frac{i}{2}\,
\varepsilon_{klj}S_j -S^{[2]}_{kl}\,.
\end{eqnarray}

Introducing now for convenience the following quantities 
\begin{eqnarray}
{[}\vec a \,\vec b \,\vec c\,]&=&
\vec a \cdot(\vec b \times \vec c\,)\,,\\
N_k&=&\frac{E_k}{M}-1=\frac{(\sqrt{s}-M)^2+K^2}{2M\sqrt{s}}\,,\\
N_q&=&\frac{E_q}{M}-1=\frac{(\sqrt{s}-M)^2-m^2}{2M\sqrt{s}}\,,\\
N &=&\frac{kq}{M^2}-N_k N_q\, \hat k \cdot \hat q \,,\\
D_k&=&\frac{k_0q}{M}+ kN_q\,\hat k \cdot \hat q\,,\\
D_q&=&\frac{q_0k}{M}+ qN_k\,\hat k \cdot \hat q\,,
\end{eqnarray}
the explicit evaluation then yields for the occurring pseudoscalars
\begin{eqnarray}
S(\epsilon,k,p,q)&=&-\sqrt{s}\,[\vec \epsilon \,\vec k \,\vec q\,]\,,\\
S(U,k,p,q)&=&-\sqrt{s}\,[\vec k \,\vec q \,\vec P]\,\chi_m^{[1]}\,,\\
S(\epsilon,U,k,p)&=&-\sqrt{s}\,[\vec \epsilon \,\vec k \,\vec P\,]
\,\chi_m^{[1]}\,,\\
S(U',k,p,q)&=&-\chi_{m'}^{[1]\,\dagger}\sqrt{s}\,
[\vec k \,\vec q \,\vec P^{\,\dagger}]\,,\\
S(U',\epsilon,k,p)&=&\chi_{m'}^{[1]\,\dagger}k\sqrt{s}\,\Big(
N_q\,[\vec \epsilon \,\hat k \,\hat q\,]
\,\hat q \cdot \vec P^{\,\dagger} 
+ [\vec \epsilon \,\hat k \,\vec P^{\,\dagger}]\Big)\,.
\end{eqnarray}
Evidently, the pseudoscalars $S(\epsilon,k,p,q)$, $S(\epsilon,U,k,p)$ and 
$S(U',\epsilon,k,p)$ lead to purely transverse currents. 
For the scalar products with the initial and final target polarization 
vectors appearing in Table \ref{tab1} and \ref{tab1b} one finds
\begin{eqnarray}
U'\cdot U&=& -\chi_{m'}^{[1]\,\dagger}\Big(\vec P^{\,\dagger}\cdot \vec P
+N_k\,\hat k \cdot \vec P^{\,\dagger}\,\hat k \cdot \vec P
+N_q\,\hat q\cdot \vec P^{\,\dagger}\, \hat q \cdot \vec P
-N\,\hat q\cdot \vec P^{\,\dagger}\,
\hat k \cdot \vec P\,\Big)\chi_m^{[1]}
\,,
\label{scalaruu}\\
U\cdot k &=& -\frac{\sqrt{s}}{M}\, \vec k \cdot \vec P\,\chi_m^{[1]}\,,
\label{scalaruk}\\
U\cdot q &=& -\Big( D_q
\,\hat k \cdot \vec P+\vec q \cdot \vec P\,\Big) \,\chi_m^{[1]}\,,
\label{scalaruq}\\
U'\cdot k &=& -\chi_{m'}^{[1]\,\dagger}\Big( D_k\,
\hat q \cdot \vec P^{\,\dagger}
+\vec k \cdot \vec P^{\,\dagger}\,\Big)\,,\label{scalarupk}\\
U'\cdot q &=& -\chi_{m'}^{[1]\,\dagger}\,\frac{\sqrt{s}}{M}\, 
\vec q \cdot \vec P^{\,\dagger}\,.\label{scalarupq}
\end{eqnarray}

Representing now the amplitudes in a reduced operator form by splitting 
off the spin functions
\begin{eqnarray}
\Omega_\alpha=\chi_{m'}^{[1]\,\dagger}\,\widetilde \Omega_\alpha
\,\chi_m^{[1]}\,,
\end{eqnarray}
one finds for the $\widetilde \Omega_\alpha$ the expressions listed 
in Table \ref{tab3}, where we have introduced the shorthand
\begin{eqnarray}
\Sigma(\vec a, \vec b\,)&=&\vec a \cdot \vec P^{\,\dagger}\, \vec b \cdot
\vec P\nonumber\\
&=&\frac{1}{3}\,\vec a\cdot\vec b +\frac{i}{2}\vec S \cdot ( \vec a \times 
 \vec b) - a_k\, S^{[2]}_{kl}\, b_l\nonumber\\
&=&\frac{1}{3}\,\vec a\cdot\vec b +\frac{i}{2}\vec S \cdot ( \vec a \times 
 \vec b) - [\vec a \times \vec b\,]^{[2]}\cdot S^{[2]}\,.\label{sigma_ab}
\end{eqnarray}
Note, that for the longitudinal operators $\widetilde \Omega_{10}$ through
$\widetilde \Omega_{13}$ the specific form of the longitudinal polarization 
vector of the virtual photon of (\ref{photpoll}) has been 
used already. As mentioned before, the operators
$\widetilde \Omega_1$ through $\widetilde \Omega_9$ are purely transverse 
and, therefore, suffice for describing photoproduction, 
whereas $\widetilde \Omega_{10}$ through $\widetilde \Omega_{13}$ 
contain only charge and longitudinal current contributions. 

It is now straightforward with the help of 
(\ref{sigma_ab}) to expand the operators $\widetilde \Omega_\alpha$ 
in terms of the nonrelativistic operators. To this 
end we first note for the various $\Sigma$-epressions in Table \ref{tab3} 
the following relations to the nonrelativistic transverse and longitudinal 
operators  ${\cal O}_{L/T,\beta}$
\begin{eqnarray}
[\vec e \,\hat k \hat q]\,\Sigma(\hat k, \hat k)&=&
\frac{1}{3}\,{\cal O}_{T,1} - {\cal O}_{T,5}\,,\\
{}[\vec e \,\hat k \hat q]\,\Sigma(\hat q, \hat q)&=&
\frac{1}{3}\,{\cal O}_{T,1} - {\cal O}_{T,7}\,,\\
{}[\vec e \,\hat k \hat q]\,\Sigma(\hat k, \hat q)&=&
\frac{1}{3}\,\hat k \cdot\hat q\,{\cal O}_{T,1} 
+\frac{i}{2}\,{\cal O}_{T,2}-{\cal O}_{T,6}\,,\\
{}[\vec e \,\hat k \hat q]\,\Sigma(\hat q, \hat k)&=&
\frac{1}{3}\,\hat k \cdot\hat q\,{\cal O}_{T,1} 
-\frac{i}{2}\,{\cal O}_{T,2}-{\cal O}_{T,6}\,,\\
\Sigma(\vec e \times \hat k, \hat k)&=&
\frac{i}{2}\,{\cal O}_{T,3} - {\cal O}_{T,8}\,,\\
\Sigma(\hat k, \vec e \times \hat k)&=&
-\frac{i}{2}\,{\cal O}_{T,3} - {\cal O}_{T,8}\,,\\
\Sigma(\vec e \times \hat k, \hat q)&=&\frac{1}{3}\,{\cal O}_{T,1}
+\frac{i}{2}\,{\cal O}_{T,4} - {\cal O}_{T,9}\,,\\
\Sigma(\hat q, \vec e \times \hat k)&=&\frac{1}{3}\,{\cal O}_{T,1}
-\frac{i}{2}\,{\cal O}_{T,4} - {\cal O}_{T,9}\,,\\
\Sigma(\hat k \times \hat q, \hat k)&=&
-\frac{i}{2}\,\hat k \cdot\hat q\,{\cal O}_{L,1} +\frac{i}{2}\,{\cal O}_{L,2}
-{\cal O}_{L,3}\,,\\
\Sigma(\hat k, \hat k \times \hat q)&=&
\frac{i}{2}\,\hat k \cdot\hat q\,{\cal O}_{L,1} -\frac{i}{2}\,{\cal O}_{L,2}
-{\cal O}_{L,3}\,,\\
\Sigma(\hat k \times \hat q, \hat q)&=&
-\frac{i}{2}\,{\cal O}_{L,1} +\frac{i}{2}\,\hat k \cdot\hat q\,{\cal O}_{L,2}
-{\cal O}_{L,4}\,,\\
\Sigma(\hat q, \hat k \times \hat q)&=&
\frac{i}{2}\,{\cal O}_{L,1} -\frac{i}{2}\,\hat k \cdot\hat q\,{\cal O}_{L,2}
-{\cal O}_{L,4}\,.
\end{eqnarray}
This then leads to the expansion of the reduced operators 
$\widetilde \Omega_\alpha$ by the ${\cal O}_{L/T,\beta}$ in the form 
\begin{eqnarray}
\widetilde \Omega_{\alpha}=
\left\{\begin{array}{ll} 
\bar g^T_\alpha\sum_{\beta=1}^9 
g^T_{\alpha,\beta}\, {\cal O}_{T,\beta} 
 & \mbox{for }\alpha=1,\dots,9\,,\\[1.ex]
\bar g^L_\alpha\sum_{\beta=1}^4 
g^L_{\alpha,\beta}\, {\cal O}_{L,\beta} 
 & \mbox{for }\alpha=10,\dots,13\,,\\
\end{array}\right.\label{omnrop} 
\end{eqnarray}
where the coefficients $\bar g^{T/L}_{\alpha}$ and $g^{T/L}_{\alpha,\beta}$ 
are listed in Tables \ref{table_gt} and \ref{table_gl}, respectively.
Corresponding relations between the invariant functions $F_\alpha(s,t,u)$ 
of (\ref{mficov}) and $G^{L/T}_\beta(s,t,u)$ of (\ref{gnonrel}) follow
\begin{eqnarray}
G^{L}_\beta(s,t,u)&=& \sum_{\alpha=10}^{13}F_\alpha(s,t,u)
\bar g^L_{\alpha}\,g^L_{\alpha,\beta}\,,\label{omegagft}\\
G^{T}_\beta(s,t,u)&=& \sum_{\alpha=1}^{9}F_\alpha(s,t,u)
\bar g^T_{\alpha}\,g^T_{\alpha,\beta}\,.\label{omegagfl}
\end{eqnarray}

With this we will close the present formal study. It will be a task for the 
future to derive explicit expressions in terms of the invariant functions 
for observables like differential cross section, target asymmetries and 
polarization components of the final target state. Furthermore, one needs to  
construct the invariant functions for specific reaction models like, 
for example, coherent pseudoscalar meson production from the deuteron 
in the impulse approximation. 
In a later stage, also rescattering and other two-body contributions can 
be treated accordingly.

\newpage

\setcounter{equation}{0}

\section*{Appendix A: Linear relations for the Levy-Civita tensor}
In this appendix we will show, that expressions containing products of three 
Levi-Civita tensors can be reduced to ones with one Levi-Civita tensor only. 
The Levi-Civita tensor is defined by 
\begin{eqnarray}
\varepsilon^{\mu\nu\rho\sigma}=
\left\{\begin{array}{rl} +1 & \mbox{if }\{\mu,\nu,\rho,\sigma\}
                              \mbox{ is an even permutation of }\{0,1,2,3\},\\
                         -1 & \mbox{if it is an odd permutation,}\\
                          0 & \mbox{otherwise.}\\
\end{array}\right. 
\end{eqnarray}
Products of two Levi-Civita tensors can be reduced to expressions containing
only the metric tensor $g^{\mu\nu}$ (see, for example, the Appendix A of 
\cite{ItZ88}). Therefore, among the products of three Levi-Civita tensors, 
the unconnected
ones, i.e., those without contractions between them, reduce to one Levy-Civita
tensor and do not lead to a linear relation. For the connected ones, 
only the following identities for contractions over one and two index 
pairs are useful for deriving linear relations
\begin{eqnarray}
\varepsilon^{\mu\nu\rho\sigma}{\varepsilon_{\sigma}}^{\mu'\nu'\rho'}&=&
det(g^{\alpha\alpha'})\,,\qquad 
 \alpha=\mu,\nu,\rho\,,\,\, \alpha'=\mu',\nu',\rho'\,,\\
\varepsilon^{\mu\nu\rho\sigma}{\varepsilon_{\rho\sigma}}^{\mu'\nu'}&=&
-2(g^{\mu\mu'}g^{\nu\nu'}-g^{\mu\nu'}g^{\nu\mu'})\,.
\end{eqnarray}
Among the various possible contractions of three Levi-Civita tensors over 
one or two index pairs between two factors, the one with two-index contractions
in both cases is also trivial
\begin{eqnarray}
\varepsilon^{\mu\nu\rho'\sigma'}\varepsilon_{\rho'\sigma'\mu'\nu'}
\varepsilon^{\mu'\nu'\rho\sigma}=-4\varepsilon^{\mu\nu\rho\sigma}\,.
\end{eqnarray}
Thus only two nontrivial cases remain. The first relation follows from 
the contraction of three Levi-Civita tensors over one index pair for the first 
product and two index pairs for the second, i.e.\
$\varepsilon^{\mu\nu\rho\sigma'}{\varepsilon_{\sigma'}}^{\lambda\nu'\rho'}
{\varepsilon_{\nu'\rho'}}^{\sigma\tau}$, resulting in 
\begin{eqnarray}
\varepsilon^{\mu\nu\rho\sigma}g^{\tau\lambda}+\varepsilon^{\nu\rho\sigma\tau}
g^{\mu\lambda}+\varepsilon^{\rho\sigma\tau\mu}g^{\nu\lambda}
+\varepsilon^{\sigma\tau\mu\nu}g^{\rho\lambda}+\varepsilon^{\tau\mu\nu\rho}
g^{\sigma\lambda}=0\,.\label{rel1a}
\end{eqnarray}
The second relation follows from the contraction over only one index pair for
each product, i.e.\ 
$\varepsilon^{\mu\nu\rho\sigma}{\varepsilon_{\sigma}}^{\alpha\beta\gamma}
{\varepsilon_{\gamma}}^{\delta\lambda\tau}$, yielding
\begin{eqnarray}
\varepsilon^{\mu\nu\rho\delta}
(g^{\alpha\lambda}g^{\beta\tau}-g^{\alpha\tau}g^{\beta\lambda})
+\varepsilon^{\mu\nu\rho\lambda}
(g^{\alpha\tau}g^{\beta\delta}-g^{\alpha\delta}g^{\beta\tau})
+\varepsilon^{\mu\nu\rho\tau}
(g^{\alpha\delta}g^{\beta\lambda}-g^{\alpha\lambda}g^{\beta\delta})&=&
\nonumber\\
\varepsilon^{\mu\delta\lambda\tau}
(g^{\alpha\nu}g^{\beta\rho}-g^{\alpha\rho}g^{\beta\nu})
+\varepsilon^{\nu\delta\lambda\tau}
(g^{\alpha\rho}g^{\beta\mu}-g^{\alpha\mu}g^{\beta\rho})
+\varepsilon^{\rho\delta\lambda\tau}
(g^{\alpha\mu}g^{\beta\nu}-g^{\alpha\nu}g^{\beta\mu})\,.
&&\label{rel2a}
\end{eqnarray}
By contraction with an appropriate number of Lorentz vectors, the first  
relation can be cast into the form
\begin{eqnarray}
S(a,b,c,d)\,e\cdot f +S(b,c,d,e)\,a\cdot f +S(c,d,e,a)\,b\cdot f
+S(d,e,a,b)\,c\cdot f+S(e,a,b,c)\,d\cdot f =0\,.
\label{rel1b}
\end{eqnarray}
Note, that this relation is totally antisymmetric in the arguments 
$(a,b,c,d,e)$. 

Defining for convenience as a shorthand 
\begin{eqnarray}
[a,b;c,d]:=(a\cdot c)(b\cdot d)-(a\cdot d)(b\cdot c)\,,\label{defklammer}
\end{eqnarray}
the second relation becomes correspondingly
\begin{eqnarray}
&&S(a,b,c,d)[e,f;g,h]+S(a,b,c,e)[f,d;g,h]+S(a,b,c,f)[d,e;g,h]=\nonumber\\
&&S(a,d,e,f)[b,c;g,h]+S(b,d,e,f)[c,a;g,h]+S(c,d,e,f)[a,b;g,h]
\,.\label{rel2b}
\end{eqnarray}
This relation is totally antisymmetric in the group of variables $(a,b,c)$, 
$(d,e,f)$, and $(g,h)$. Furthermore, it is symmetric under the common 
interchange $(a,b,c)\leftrightarrow (d,e,f)$. This property will be useful 
in the following application. From these relations follows furthermore, that
higher order, odd power products of the Levi-Civita tensor can always be 
reduced to linear expressions.

\setcounter{equation}{0}

\section*{Appendix B: Linear relations between invariant amplitudes}

We will first exploit the relation in (\ref{rel1b}). In view of 
its symmetry property, one can distinguish two cases with respect to 
the three polarization vectors $(U',U,\epsilon)$. 

(i) All three belong to the group of variables $(a,b,c,d)$. In this case 
one finds
\begin{eqnarray}
S(U',\epsilon,U,x)\,y\cdot z +U'\cdot z\,S(\epsilon,U,x,y) 
-S(U',U,x,y)\,\epsilon\cdot z&&\nonumber\\
+S(U',\epsilon,x,y)\,U\cdot z-S(U',\epsilon,U,y)\,x\cdot z&=&0\,,
\end{eqnarray}
which results in the following relation between the amplitudes of Table
\ref{tab1}
\begin{eqnarray}
\Omega_d(x,y,z)&=&\Omega_a(x)\,y\cdot z-\Omega_a(y)\,x\cdot z 
+\Omega_b(x,y,z)+\Omega_c(x,y,z)\,.\label{rel11}
\end{eqnarray}
It allows to eliminate completely $\Omega_d(x,y,z)$ in favour of $\Omega_a$, 
$\Omega_b$ and $\Omega_c$. 

(ii) One of the polarization vectors is identified with $f$ in (\ref{rel1b}). 
This yields three relations. First setting $f$ equal to $U$ and $U'$, one 
obtains
\begin{eqnarray}
\Omega_b(x,y,z)+\Omega_b(z,x,y)+\Omega_b(y,z,x)+
[\Omega_e(U',\epsilon,U)-\Omega_e(\epsilon,U',U)]\varepsilon(x,y,z)\,,
\label{rel12}\\
\Omega_c(x,y,z)+\Omega_c(z,x,y)+\Omega_c(y,z,x)=
[\Omega_e(\epsilon,U',U)-\Omega_e(U,U',\epsilon)]\varepsilon(x,y,z)\,,
\label{rel13}
\end{eqnarray}
where we have defined 
\begin{eqnarray}
\varepsilon(x,y,z)=
\left\{\begin{array}{rl} +1 & \mbox{if }\{x,y,z\}
                              \mbox{ is an even permutation of }\{k,p,q\},\\
                         -1 & \mbox{if it is an odd permutation,}\\
                          0 & \mbox{otherwise.}\\
\end{array}\right. 
\end{eqnarray}
These two relations serve to eliminate $\Omega_b(p,q,k)$ and 
$\Omega_c(p,q,k)$. 

Finally, the last relation in this group which follows by setting $f=\epsilon$
\begin{eqnarray}
\Omega_d(x,y,z)+\Omega_d(y,z,x)+\Omega_d(z,x,y)=
\left[\Omega_e(U',U,\epsilon)-\Omega_e(U,U',\epsilon)
\right]\varepsilon(x,y,z)\,,
\end{eqnarray}
does not constitute an additional constraint. It is fulfilled
identically by use of the previous relations (\ref{rel11}) through 
(\ref{rel13}).

The exploitation of the second relation (\ref{rel2b}) is more involved. 
One may distinguish five groups according to the location of the polarization 
vectors among the arguments $(a,b,c,d,e,f,g,h)$. They are listed in Table 
\ref{tab2}. 

Before evaluating these various groups, several shorthand notations will 
be introduced 
\begin{eqnarray}
Q_a(x;y,z;u,v):= [y,z;u,v]\Omega_a(x)\,,
\end{eqnarray}
which is antisymmetric in $(y,z)$ and in $(u,v)$ and symmetric under the 
interchange $(y,z)\leftrightarrow (u,v)$. Furthermore, for the amplitudes of 
type $\Omega_\alpha$ we define for $\alpha\in \{b,c\}$
\begin{eqnarray}
Q_\alpha(x,y,z;u,v)&:=& u\cdot x\, \Omega_\alpha(y,z,v) +
u\cdot y\, \Omega_\alpha(z,x,v)+u\cdot z\, \Omega_\alpha(x,y,v)\,,
\label{defqbc}
\end{eqnarray}
and for $\beta \in \{f,g,h\}$
\begin{eqnarray}
Q_\beta(x,y,z;u,v)&:=& \varepsilon(x,y,z)\Omega_\beta(u,v)\,,
\end{eqnarray}
which are both totally antisymmetric in $(x,y,z)$. 
And finally we define
\begin{eqnarray}
{\cal O}_a(x,y,z;u,v)&:=& Q_a(x;y,z;u,v)+Q_a(y;z,x;u,v)+Q_a(z;x,y;u,v)\,,\\
{\cal O}_\alpha(x,y,z;u,v)&:=& Q_\alpha(x,y,z;u,v)-Q_\alpha(x,y,z;v,u)\,,
\end{eqnarray}
for amplitudes of type $\alpha\in \{b,c,f,g,h\}$, 
which are again all antisymmetric in $(x,y,z)$ and in $(u,v)$.

Now we will consider the relations which correspond to the various groups 
in Tab.~\ref{tab2}.
\newline

\noindent
(1) The first group leads straightforwardly to
\begin{eqnarray}
{\cal O}_a(x,y,z;u,v)+{\cal O}_h(x,y,z;u,v)=
{\cal O}_f(x,y,z;u,v)+{\cal O}_g(x,y,z;u,v)\,.
\end{eqnarray}

\noindent
(2) The second group leads to three relations
\begin{eqnarray}
{\cal O}_a(x,y,z;u,v)&=&
{\cal O}_b(x,y,z;u,v)+{\cal O}_g(x,y,z;u,v)\,,\\
{\cal O}_a(x,y,z;u,v)&=&
{\cal O}_c(x,y,z;u,v)+{\cal O}_f(x,y,z;u,v)\,,\\
{\cal O}_a(x,y,z;u,v)&=&
{\cal O}_b(x,y,z;u,v)+{\cal O}_c(x,y,z;u,v)+{\cal O}_h(x,y,z;u,v)\,.
\label{rel22c}
\end{eqnarray}
From these one can also derive the relation of group (1).
\newline

\noindent
(3) For the third group one finds, using the identity II of Appendix C,
\begin{eqnarray}
Q_f(x,y,z;u,v)-Q_h(x,y,z;u,v)&=&Q_b(x,y,z;u,v)\,,\label{rel23a}\\
Q_g(x,y,z;u,v)-Q_h(x,y,z;u,v)&=&-Q_c(x,y,z;v,u)\,,\label{rel23b}\\
Q_g(x,y,z;v,u)-Q_f(x,y,z;u,v)&=&-{\cal O}_a(x,y,z;u,v)-Q_b(x,y,z;v,u)
\nonumber\\
&&-Q_c(x,y,z;v,u)\,.\label{rel23c}
\end{eqnarray}
It is easy to see, that these relations contain the foregoing ones of the 
groups (1) and (2). 
\newline

\noindent
(4) For this group one obtains
\begin{eqnarray}
Q_f(x,y,z;v,u)-Q_h(x,y,u;v,z)&=&Q_b(y,z,u;v,x)-Q_b(x,z,u;v,y)\nonumber\\
                         &+&Q_h(y,z,u;v,x)-Q_h(x,z,u;v,y)\,,\\
Q_g(x,y,z;u,v)-Q_g(u,z,x;y,v)&=&Q_c(u,y,x;v,z)-Q_c(y,z,u;v,x)\nonumber\\
                         &+&Q_h(y,z,u;x,v)-Q_h(u,y,x;z,v)\,,\\
Q_g(x,y,z;v,u)-Q_g(x,y,u;v,z)&+&Q_f(x,y,z;z,v)-Q_f(x,y,u;z,v)=\nonumber\\
{\cal O}_a(x,y,v;z,u)&-&{\cal O}_a(z,u,v;x,y)+Q_b(x,y,u;v,z)
\nonumber\\
-Q_b(x,y,z;v,u)&+&Q_c(x,y,u;v,z)-Q_c(x,y,z;v,u)\,.
\end{eqnarray}
In these relations, originally also $Q_e$ appears formally, but its 
contribution drops out due to the fact that
\begin{eqnarray}
v\cdot [x\, \varepsilon(y,z,u) - y\,\varepsilon(z,u,x)+z\,\varepsilon(u,x,y)
+u\,\varepsilon(x,y,z)] = 0
\end{eqnarray}
according to the identity (\ref{ident1}) in Appendix C.
Also these relations are contained in the ones of group (3).
\newline

\noindent
(5) Finally, for the last group one has
\begin{eqnarray}
{\cal O}_\alpha(x,y,z;u,v)+{\cal O}_\alpha(x,y,u;v,z)+
{\cal O}_\alpha(x,y,v;z,u)={\cal O}_\alpha(z,u,v;x,y)
\end{eqnarray}
for $\alpha\in \{f,g,h\}$, which are identities because of (\ref{ident2}).
This completes the search for all possible relations among the various 
amplitudes listed in Table \ref{tab0}.

Thus, of all relations of the various groups, only the ones of group (3)
constitute new information, which allow to further eliminate some of the 
amplitudes $\Omega_\alpha$. To this end we set first in (\ref{rel23a}) and 
(\ref{rel23b}) $(x,y,z)=(k,p,q)$ and then $(u,v)=(x,y)$ and find
\begin{eqnarray}
\Omega_f(x,y)=\Omega_h(x,y)+Q_b(k,p,q;x,y)\,,\\
\Omega_g(x,y)=\Omega_h(x,y)-Q_c(k,p,q;y,x)\,,
\end{eqnarray}
by which $\Omega_f$ and $\Omega_g$ can be eliminated. Inserting these 
expressions into (\ref{rel23c}) one finds 
\begin{eqnarray}
{\cal O}_b(k,p,q;x,y)+{\cal O}_c(k,p,q;x,y)=
{\cal O}_a(k,p,q;x,y)-{\cal O}_h(k,p,q;x,y)\,,
\label{rel22cs}
\end{eqnarray}
which corresponds to (\ref{rel22c}). Because of the asymmetry in $(x,y)$,
one has three different cases which serve to eliminate the remaining 
amplitudes $\Omega_b(p,q;q)$,
$\Omega_c(p,q;q)$ and $\Omega_e(\epsilon,U'U)$. To this end we define
for $\alpha\in \{b,c\}$
\begin{eqnarray}
\widetilde Q_\alpha(x,y)&:=&p\cdot x\,\Omega_\alpha(q,k,y)+q\cdot x\,
\Omega_\alpha(k,p,y)
\,,\\
\widetilde {\cal O}(x,y)&:=& {\cal O}_h(k,p,q;x,y)+\widetilde Q_b(x,y)
-\widetilde Q_b(y,x)
+\widetilde Q_c(x,y)-\widetilde Q_c(y,x)\,,\\
{\cal P}(x,y)&:=&{\cal O}_a(k,p,q;x,y)-\widetilde {\cal O}(x,y)\,.
\end{eqnarray}
Note that $\widetilde {\cal O}(x,y)$ contains gauge invariant amplitudes 
only but not $\Omega_a(k)$. Then (\ref{rel22cs}) becomes 
\begin{eqnarray}
k\cdot x\,(\Omega_b(p,q,y)+\Omega_c(p,q,y)) 
- k\cdot y\,[\Omega_b(p,q,x)+\Omega_c(p,q,y)]={\cal P}(x,y)\,,
\end{eqnarray}
where ${\cal P}(x,y)$ does not contain $\Omega_b(p,q;x)$ and $\Omega_c(p,q;x)$.
Taking first $(x,y)=(k,p)$, one obtains a relation for $\Omega_c(p,q,q)$
\begin{eqnarray}
k^2\,\Omega_c(p,q,q)&=& k^2\,\Omega_c(p,q,k)+
 k\cdot p\,[\Omega_b(p,q,k)+\Omega_c(p,q,k)]+
{\cal P}(k,p)\,.\label{relqc}
\end{eqnarray}
Then setting $(x,y)=(k,q)$ yields $\Omega_b(p,q,q)$
\begin{eqnarray}
k^2\,\Omega_b(p,q,q)&=& -k^2\,\Omega_c(p,q,q)+
 k\cdot q\,[\Omega_b(p,q,k)+\Omega_c(p,q,k)]+
{\cal P}(k,q)\,,\nonumber\\
&=&(k\cdot q-k\cdot p)\Omega_b(p,q,k)+(k\cdot q-k\cdot p-k^2)\Omega_c(p,q,k)
\nonumber\\
&&+{\cal P}(k,q)-{\cal P}(k,p)\,.\label{relqb}
\end{eqnarray}
Finally for $(x,y)=(p,q)$, using the previous two equations,
one gets a relation 
\begin{eqnarray}
k\cdot p\,{\cal P}(k,q)-k\cdot q\,{\cal P}(k,p)-k^2\,{\cal P}(p,q)=0\,.
\label{omegae}
\end{eqnarray}
which allows to eliminate $\Omega_a(k)$. To this end we rewrite (\ref{omegae})
yielding
\begin{eqnarray}
k\cdot p\,{\cal O}_a(k,q)-k\cdot q\,{\cal O}_a(k,p)-k^2\,{\cal O}_a(p,q)=
k\cdot p\,\widetilde {\cal O}(k,q)-k\cdot q\,\widetilde {\cal O}(k,p)-k^2\,
\widetilde {\cal O}(p,q)\,,
\label{omegae1}
\end{eqnarray}
where the rhs is gauge invariant and does not contain $\Omega_a(k)$. 
Thus the non-gauge invariant amplitudes $\Omega_a(p)$ and $\Omega_a(q)$ 
have to drop out also on the lhs, and indeed one finds
\begin{eqnarray}
(k^2p^2q^2 -k^2(p\cdot q)^2 -p^2(k\cdot q)^2 -q^2 (k\cdot p)^2+2\,
k\cdot p\,k\cdot q\,p\cdot q)\Omega_a(k)=
\nonumber\\
k\cdot p\,\widetilde {\cal O}(k,q)-k\cdot q\,\widetilde {\cal O}(k,p)-k^2\,
\widetilde {\cal O}(p,q)\,.\label{omegae2}
\end{eqnarray}
This is the required relation in order to eliminate one of the 14 
gauge invariant amplitudes for which we have chosen $\Omega_a(k)$.

\setcounter{equation}{0}

\section*{Appendix C: Two identities}

In this appendix we note two useful identities. 

(I) Given a function $f(x,y,z;u)$ which is totally antisymmetric in 
$(x,y,z)$, and for which the arguments refer to only three independent 
variables, then the following identity holds
\begin{eqnarray}
f(x,y,z;u)-f(y,z,u;x)+f(z,u,x;y)-f(u,x,y;z)=0\,,\label{ident1}
\end{eqnarray}
the proof of which is straightforward and almost trivial, for example, 
by assuming $(x,y,z)$ to be independent and then setting $u=x$. Because 
of the asymmetry, this then holds also for $u=y,z$.

(II) Given a function $g(x,y,z;u,v)$ which is totally antisymmetric in 
$(x,y,z)$ and in $(u,v)$, and for which all arguments refer to only three 
independent variables, then the following identity holds
\begin{eqnarray}
g(x,y,z;u,v)+g(x,y,u;v,z)+g(x,y,v;z,u)=g(z,u,v;x,y)\,,\label{ident2}
\end{eqnarray}
Since both sides are totally antisymmetric in $(z,u,v)$, it is sufficient to
assume $(z,u,v)$ as independent. Then $(x,y)$ have to coincide with two of 
them and then it is again straightforward to 
show the validity of (\ref{ident2}).

\setcounter{equation}{0}

\newpage
\begin{table}[h]
\caption{Set of basic types of invariant amplitudes with 
$x,y,z\in \{k,p,q\}$ and $a,b,c \in \{U',\epsilon, U\}$.} 
\begin{center}

\begin{tabular}{|c|l|}\hline
  notation & explicit form\\
\hline
 $\Omega_a(x)$ & $S(U',\epsilon, U, x)$\\
 $\Omega_b(x, y, z)$ & $S(U',\epsilon, x, y ) U\cdot z$\\
 $\Omega_c(x, y, z)$ & $U'\cdot z\,S(\epsilon, U, x, y)$\\
 $\Omega_d(x, y, z)$ & $S(U', U, x, y )\epsilon \cdot z$\\
 $\Omega_e(a, b, c)$ & $S(a, k,p,q)\,b\cdot c$\\
 $\Omega_f(x, y)$ & $S(U',k, p, q)\,\epsilon \cdot x \,U\cdot y$\\
 $\Omega_g(x, y)$ & $U'\cdot x \,S(U, k, p, q)\,\epsilon \cdot y$\\
 $\Omega_h(x, y)$ & $U'\cdot x \, S(\epsilon,k, p, q)\,U\cdot y$\\
\hline
\end{tabular}
\end{center}
\label{tab0}
\end{table}

\begin{table}[h]
\caption{Set of independent gauge invariant amplitudes $\Omega_\alpha$.}
\begin{center}

\begin{tabular}{|r|c|l|}\hline
$\alpha$ & notation & explicit form\\
\hline
1 & $\Omega_e(\epsilon,U',U)$ & $S(\epsilon,k,p,q)U'\cdot U$\\
2 & $\Omega_h(k,k)$ & $U'\cdot k\,S(\epsilon, k,p,q )U\cdot k$\\
3 & $\Omega_h(k,q)$ & $U'\cdot k\,S(\epsilon, k,p,q )U\cdot q$\\
4 & $\Omega_h(q,k)$ & $U'\cdot q\,S(\epsilon, k,p,q )U\cdot k$\\
5 & $\Omega_h(q,q)$ & $U'\cdot q\,S(\epsilon, k,p,q )U\cdot q$\\
6 & $\Omega_b(k,p,k)$ & $S(U',\epsilon, k,p ) U\cdot k$\\
7 & $\Omega_b(k,p,q)$ & $S(U',\epsilon, k,p ) U\cdot q$\\
8 & $\Omega_c(k,p,k)$ & $U'\cdot k\,S(\epsilon,U, k,p )$\\
9 & $\Omega_c(k,p,q)$ & $U'\cdot q\,S(\epsilon,U, k,p )$\\
\hline
10 & $\Omega_b(k,q,k)$ & $S(U',\epsilon, k,q ) U\cdot k$\\
11 & $\Omega_b(k,q,q)$ & $S(U',\epsilon, k,q ) U\cdot q$\\
12 & $\Omega_c(k,q,k)$ & $U'\cdot k\,S(\epsilon,U, k,q )$\\
13 & $\Omega_c(k,q,q)$ & $U'\cdot q\,S(\epsilon,U, k,q )$\\
\hline
\end{tabular}
\end{center}
\label{tab1}
\end{table}

\begin{table}[h]
\caption{Set of independent non-gauge invariant amplitudes.}
\begin{center}

\begin{tabular}{|r|c|l|}\hline
$\alpha$ & notation & explicit form\\
\hline
14 & $\Omega_a(p)$ & $S(U',\epsilon, U, p)$\\
15 & $\Omega_a(q)$ & $S(U',\epsilon, U, q)$\\
16 & $\Omega_e(U', \epsilon, U)$ & $S(U',k,p,q)\,\epsilon\cdot U$\\
17 & $\Omega_e(U,U', \epsilon)$ & $U'\cdot\epsilon\, S(U,k,p,q)$\\
\hline
\end{tabular}
\end{center}
\label{tab1a}
\end{table}

\begin{table}[h]
\caption{Equivalent set of independent gauge invariant longitudinal
amplitudes for electroproduction.}
\begin{center}

\begin{tabular}{|r|l|l|}\hline
$\alpha$ & notation & explicit form\\
\hline
10 & $k\cdot p\, \Omega_f(k,k) - k^2\,\Omega_f(p,k)$ 
& $S(U',k,p,q )\,U \cdot k\,
(k\cdot p\,\, k\cdot \epsilon - k^2\,p\cdot \epsilon)$\\
11 & $k\cdot p\, \Omega_f(k,q) - k^2\,\Omega_f(p,q)$ 
& $S(U',k,p,q )\,U \cdot q\,
(k\cdot p\,\, k\cdot \epsilon - k^2\,p\cdot \epsilon)$\\
12 & $k\cdot p\, \Omega_g(k,k) - k^2\,\Omega_g(k,p)$ & 
$U'\cdot k\,S(U, k,p,q )
(k\cdot p\,\, k\cdot \epsilon - k^2\,p\cdot \epsilon)$\\
13 & $k\cdot p\, \Omega_g(q,k) - k^2\,\Omega_g(q,p)$ & 
$U'\cdot q\,S(U, k,p,q )
(k\cdot p\,\, k\cdot \epsilon - k^2\,p\cdot \epsilon)$\\
\hline
\end{tabular}
\end{center}
\label{tab1b}
\end{table}

\begin{table}[h]
\caption{Helicity representation of invariant amplitudes.}
\begin{center}

\begin{tabular}{|r|l|}\hline
$\alpha$ & $\Omega_{\alpha,\,\lambda' \lambda_\gamma \lambda}$\\
\hline
&\\[-2.ex]
1 & $\frac{i}{M^2}\, \lambda_\gamma\, kq\sqrt{s}\,
      \Big[p'p\,\delta_{\lambda' 0}\delta_{\lambda 0}
      - E'(\lambda')E(\lambda)\, d^1_{\lambda \lambda'}(\theta)\Big]
      \,d^1_{-\lambda_\gamma 0}(\theta)$\\[1.ex]
2 & $\frac{i}{M^2}\, \lambda_\gamma\, k^2qs\,
     \Big[qk_0\,\delta_{\lambda' 0}+kE'(\lambda')\,d^1_{0\lambda'}(\theta)\Big]
      \,d^1_{-\lambda_\gamma 0}(\theta)\delta_{\lambda 0}$\\[1.ex]
3 & $\frac{i}{M^2}\, \lambda_\gamma\, kq\sqrt{s}\,
     \Big[qk_0\,\delta_{\lambda' 0}+kE'(\lambda')\,d^1_{0\lambda'}(\theta)\Big]
      \Big[kq_0\,\delta_{\lambda 0}+qE(\lambda)\,d^1_{\lambda 0}(\theta)\Big]
      \,d^1_{-\lambda_\gamma 0}(\theta)$\\[1.ex]
4 & $\frac{i}{M^2}\, \lambda_\gamma\, k^2q^2s^{3/2}
      \,\delta_{\lambda' 0}\delta_{\lambda 0}
      \,d^1_{-\lambda_\gamma 0}(\theta)$\\[1.ex]
5 & $\frac{i}{M^2}\, \lambda_\gamma\, kq^2s\,\delta_{\lambda' 0}\,
      \Big[kq_0\,\delta_{\lambda 0}+qE(\lambda)\,d^1_{\lambda 0}(\theta)\Big]
      \,d^1_{-\lambda_\gamma 0}(\theta)$\\[1.ex]
6 & $-\frac{i}{M^2}\, (-)^{\lambda'}\lambda_\gamma\, E'(\lambda')k^2s
      \,\delta_{\lambda 0}\,d^1_{\lambda_\gamma-\lambda'}(\theta)$\\[1.ex]
7 & $-\frac{i}{M^2}\, (-)^{\lambda'}\lambda_\gamma\, E'(\lambda')k\sqrt{s}
     \,\Big[kq_0\,\delta_{\lambda 0}+qE(\lambda)\,d^1_{\lambda 0}(\theta)\Big]
     \,d^1_{\lambda_\gamma-\lambda'}(\theta)$\\[1.ex]
8 & $\frac{i}{M^2}\,\lambda\, E(\lambda)k\sqrt{s}\,
     \Big[qk_0\,\delta_{\lambda' 0}+kE'(\lambda')\,d^1_{0\lambda'}(\theta)\Big]
      \,\delta_{\lambda \lambda_\gamma}$\\[1.ex]
9 & $\frac{i}{M^2}\,\lambda\, E(\lambda)kqs\,\delta_{\lambda' 0}
     \,\delta_{\lambda_\gamma \lambda}$\\[1.ex]
\hline
&\\[-2.ex]
10 & $\frac{i}{M}\, \lambda'\,\sqrt{K^2}\, k^3qs^{3/2}\,
     d^1_{\lambda'0}(\theta)\,\delta_{ \lambda_\gamma 0}
     \,\delta_{\lambda 0}$\\[1.ex]
11 & $\frac{i}{M}\, \lambda'\,\sqrt{K^2}\, k^2qs\,d^1_{\lambda'0}(\theta)
     \,\delta_{ \lambda_\gamma 0}\,
     \Big[kq_0\,\delta_{\lambda 0}+qE(\lambda)\,d^1_{\lambda 0}(\theta)\Big]$
\\[1.ex]
12 & $-\frac{i}{M}\, \lambda\,\sqrt{K^2}\, k^2qs\,
     \Big[qk_0\,\delta_{\lambda' 0}+kE'(\lambda')\,d^1_{0\lambda'}(\theta)\Big]
      \,\delta_{ \lambda_\gamma 0}\,d^1_{\lambda0}(\theta) $\\[1.ex]
13 & $-\frac{i}{M} \lambda\,\sqrt{K^2}\, k^2q^2s^{3/2}\,
     \delta_{\lambda'0}\,\delta_{ \lambda_\gamma 0}
     \,d^1_{\lambda 0}(\theta)$\\[1.ex]
\hline
\end{tabular}
\end{center}
\label{tabhelicity}
\end{table}

\begin{table}[h]
\caption{Basic set of independent, nonrelativistic transverse and 
longitudinal operators.}
\begin{center}

\begin{tabular}{|c|l|l|}\hline
 $\beta$ & ${\cal O}_{T,\beta} $ & ${\cal O}_{L,\beta}$ \\  
\hline
1 & $\vec \epsilon\cdot (\hat k \times \hat q\,) $ & 
$\hat k \cdot \vec S $\\
2 & $\vec \epsilon\cdot (\hat k \times \hat q\,)\,
(\hat k \times \hat q\,)\cdot \vec S  $ &
$\hat q \cdot \vec S$ \\
3 & $\vec \epsilon\cdot (\hat k \times (\hat k \times \vec S\,))$ & 
$[(\hat k\times \hat q\,)\times \hat k\,]^{[2]}\cdot S^{[2]}  $ \\
4 & $\vec \epsilon\cdot (\hat k \times (\hat q \times \vec S\,))$ &
$[(\hat k\times \hat q\,)\times \hat q\,]^{[2]}\cdot S^{[2]} $\\
5 & $\vec \epsilon\cdot (\hat k \times \hat q\,)\,\hat k^{[2]}\cdot S^{[2]}$ 
& \\
6 & $\vec \epsilon\cdot (\hat k \times \hat q\,)\,[\hat k\times \hat q]^{[2]}
\cdot S^{[2]}  $ & \\
7 & $\vec \epsilon\cdot (\hat k \times \hat q\,)\, \hat q^{[2]}\cdot S^{[2]}  $ & \\
8 & $\vec \epsilon\cdot (\hat k \times [\hat k\times S^{[2]}]^{[1]})$ & \\
9 & $\vec \epsilon\cdot (\hat k \times [\hat q\times S^{[2]}]^{[1]})$ & \\
\hline
\end{tabular}
\end{center}
\label{tabkqls}
\end{table}

\begin{table}[h]
\caption{Reduced gauge invariant amplitudes $\widetilde \Omega_\alpha$.}
\begin{center}

\begin{tabular}{|c|l|}\hline
&\\[-2.ex]
$\widetilde \Omega_1$ & $kq\sqrt{s}\,
[\vec \epsilon \,\hat k \,\hat q\,]\,\Big[1 
+N_k\,\Sigma(\hat k , \hat k\,)
+N_q\,\Sigma(\hat q , \hat q\,)
-N\,\Sigma( \hat q,\hat k \,)
\Big]$
\\[1.ex]

$\widetilde \Omega_2$ & $-\frac{1}{M}\,k^2qs\,
[\vec \epsilon \,\hat k \,\hat q\,]\,
\Big[k\Sigma(\hat k , \hat k\,)+D_k\,\Sigma(\hat q , \hat k\,)\Big]$
\\[1.ex]

$\widetilde \Omega_3$ & $-kq\sqrt{s}\,
[\vec \epsilon \,\hat k \,\hat q\,]
\,\Big[kq\Sigma(\hat k , \hat q\,)
+qD_k\,\Sigma(\hat q , \hat q\,)
+kD_q\,\Sigma(\hat k , \hat k\,)+
D_kD_q\,\Sigma(\hat q , \hat k\,)\Big]$
\\[1.ex]

$\widetilde \Omega_4$ & $-\frac{1}{M^2}\,k^2q^2s^{3/2}\,
[\vec \epsilon \,\hat k \,\hat q\,]\Sigma(\hat q , \hat k\,)$
\\[1.ex]

$\widetilde \Omega_5$ & $-\frac{1}{M}\,kq^2s\,
[\vec \epsilon \,\hat k \,\hat q\,]\,
\Big[q\Sigma(\hat q , \hat q\,)+D_q\,\Sigma(\hat q , \hat k\,)\Big]$
\\[1.ex]

$\widetilde \Omega_6$ & $-\frac{1}{M}\,k^2s\,
\Big[\Sigma(\vec \epsilon \times \hat k, \hat k\,)
+N_q\,[\vec \epsilon \,\hat k \,\hat q\,]\,
\Sigma(\hat q , \hat k\,)\Big] $\\[1.ex]

$\widetilde \Omega_7$ & $-k\sqrt{s}\,\Big[
q\Sigma(\vec \epsilon \times \hat k , \hat q\,)
+D_q\,\Sigma(\vec \epsilon \times \hat k , \hat k\,)
+N_q\,[\vec \epsilon \,\hat k \,\hat q\,]\Big(
q\Sigma(\hat q , \hat q\,)+
D_q\,\Sigma(\hat q , \hat k\,)\Big)\Big]
$\\[1.ex]

$\widetilde \Omega_8$ & $k\sqrt{s}\,\Big[k
\Sigma(\hat k,\vec \epsilon \times \hat k\,)
+D_k\,\Sigma(\hat q,\vec \epsilon \times \hat k\,)\Big]
$\\[1.ex]

$\widetilde \Omega_9$ & $\frac{1}{M}\,kqs\,
\Sigma(\hat q,\vec \epsilon \times \hat k\,)
$\\[1.ex]
\hline
&\\[-2.ex]

$\widetilde \Omega_{10}$ & $
\frac{1}{M}\,\sqrt{K^2}\,k^3q s^{3/2}\,\Sigma(\hat k \times \hat q,\hat k\,)
$\\[1.ex]

$\widetilde \Omega_{11}$ & $\sqrt{K^2}\,k^2qs
\,\Big[q\Sigma(\hat k \times \hat q,\hat q\,)
+D_q\,\Sigma(\hat k \times \hat q,\hat k\,)\Big]$\\[1.ex]

$\widetilde \Omega_{12}$ & $\sqrt{K^2}\, k^2qs
\,\Big[k\Sigma(\hat k,\hat k \times \hat q\,)
+D_k\,\Sigma(\hat q,\hat k \times \hat q\,)\Big]$\\[1.ex]

$\widetilde \Omega_{13}$ & $\frac{1}{M}\,\sqrt{K^2}\, 
k^2q^2s^{3/2}\,\Sigma(\hat q,\hat k \times \hat q\,)$
\\[1.ex]
\hline
\end{tabular}
\end{center}
\label{tab3}
\end{table}

\begin{table}[h]
\caption{Coefficients for the transformation of the reduced transverse 
operators $\widetilde \Omega_\alpha$ $(\alpha=1,\dots,9)$ to the 
nonrelativistic operators ${\cal O}_{T,\beta}$
according to (\protect{\ref{omnrop}}).}
\begin{center}

\begin{tabular}{|c|c|c|c|c|c|}\hline
&&&&&\\[-2.ex]
$\alpha$ & $\bar g_\alpha^T$ & $g^T_{\alpha,1}$ & $g^T_{\alpha,2}$ & 
$g^T_{\alpha,3}$ & $g^T_{\alpha,4}$ 
\\[1.ex]
\hline
&&&&&\\[-2.ex]
1& $kq\sqrt{s}$  & $1 + \frac{1}{3}(N_k + N_q -N \hat k \cdot \hat q)$ &
$\frac{i}{2}N$ & & 
\\[1.ex]
2& $\frac{1}{M}\,k^2qs$ & $-\frac{1}{3}(k+D_k\hat k \cdot \hat q)$ &
$\frac{i}{2}D_k$ & & 
\\[1.ex]
3& $kq\sqrt{s}$ & $-\frac{1}{3}(kD_q+qD_k+(kq+D_kD_q) \hat k \cdot \hat q)$ 
& $-\frac{i}{2}(kq-D_kD_q)$ & & 
\\[1.ex]
4& $\frac{1}{M^2}\,k^2q^2s^{3/2}$ & $-\frac{1}{3}\hat k \cdot\hat q$ &
$\frac{i}{2}$ & & 
\\[1.ex]
5& $\frac{1}{M}kq^2s$ & $-\frac{1}{3}(q+D_q\hat k \cdot \hat q)$ & 
$\frac{i}{2}D_q$ & & 
\\[1.ex]
6& $\frac{1}{M}k^2s$ & $-\frac{1}{3}N_q\hat k \cdot\hat q$ & 
$\frac{i}{2}N_q$ & & 
\\[1.ex]
7& $k\sqrt{s}\,$ & $-\frac{1}{3}(q+N_q(q+D_q\hat k \cdot\hat q))$ & 
$\frac{i}{2}D_qN_q$ & $-\frac{i}{2}D_q$ & $-\frac{i}{2}q$ 
\\[1.ex]
8& $k\sqrt{s}\,$ & $\frac{1}{3}D_k$ & & $-\frac{i}{2}k$ & 
$-\frac{i}{2}D_k$  
\\[1.ex]
9& $\frac{1}{M}\,kqs$ & $\frac{1}{3}$ & $-\frac{i}{2}$ & & 
\\[1.ex]
\hline
\hline
&&&&&\\[-2.ex]
$\alpha$ & $g^T_{\alpha,5}$ & $g^T_{\alpha,6}$ & 
$g^T_{\alpha,7}$ & $g^T_{\alpha,8}$ & $g^T_{\alpha,9}$ 
\\[1.ex]
\hline
&&&&&\\[-2.ex]
1& $-N_k$ &$N$ &$-N_q$ & &
\\[1.ex]
2&  $k$ & $D_k$ & & &
\\[1.ex]
3&  $kD_q$ & $kq+D_kD_q$ & $qD_k$ & & 
\\[1.ex]
4&  & $1$ & & &
\\[1.ex]
5&  & $D_q$ & $q$ & &
\\[1.ex]
6&  & $N_q$ & $-\frac{i}{2}$ & $1$ & 
\\[1.ex]
7&  & 
$D_qN_q$ & $qN_q$ & $D_q$ & $q$
\\[1.ex]
8& & & & $-k$ & $-D_k$
\\[1.ex]
9& & & & & $-1$
\\[1.ex]
\hline
\end{tabular}
\end{center}
\label{table_gt}
\end{table}

\begin{table}[h]
\caption{Coefficients for the transformation of the reduced longitudinal 
operators $\widetilde \Omega_\alpha$ $(\alpha=10,\dots,13)$ to the 
nonrelativistic operators ${\cal O}_{L,\beta}$
according to (\protect{\ref{omnrop}}).}
\begin{center}

\begin{tabular}{|c|c|c|c|c|c|}\hline
&&&&&\\[-2.ex]
$\alpha$ & $\bar g_\alpha^L$ & $g^L_{\alpha,1}$ & $g^L_{\alpha,2}$ & 
$g^L_{\alpha,3}$ & $g^L_{\alpha,4}$ 
\\[1.ex]
\hline
&&&&&\\[-2.ex]
10& $\frac{1}{M}\,\sqrt{K^2}\,k^3qs^{3/2}$ & 
$-\frac{i}{2}\hat k \cdot \hat q$ & $\frac{i}{2}$ & $-1$ &
\\[1.ex]
11& $\sqrt{K^2} \,k^2 qs$ & 
$-\frac{i}{2}(q+D_q\hat k \cdot \hat q)$ & 
$\frac{i}{2}(q\,\hat k \cdot \hat q+D_q)$ & 
$-D_q$ & $-q$ 
\\[1.ex]
12& $\sqrt{K^2}\, k^2 qs$ & 
$\frac{i}{2}(k\,\hat k \cdot \hat q+D_k)$ & 
$-\frac{i}{2}(k+D_k\hat k \cdot \hat q)$ & $-k$ & $-D_k$ 
\\[1.ex]
13& $\frac{1}{M}\,\sqrt{K^2}\,k^2q^2s^{3/2}$ & $\frac{i}{2}$ & 
$-\frac{i}{2}\hat k \cdot \hat q$ & & $-1$ 
\\[1.ex]
\hline
\end{tabular}
\end{center}
\label{table_gl}
\end{table}

\setcounter{table}{0}

\begin{table}[h]
\caption{Listing of various groups of relations following from 
(\protect{\ref{rel2b}}).}
\begin{center}

\begin{tabular}{|c|c|c|c|}\hline
 &$a,b,c$ & $d,e,f$ &$g,h$  \\
\hline
(1) &$U',\epsilon,U$ & & \\
\hline
 &$U',\epsilon$ &$U$ & \\
(2) &$\epsilon,U$ &$U'$ & \\
 &$U',U$ &$\epsilon$ & \\
\hline
 &$U',\epsilon$ & &$U$ \\
(3) &$\epsilon,U$ & &$U'$ \\
 &$U',U$ & &$\epsilon$ \\
\hline
 &$U'$ &$\epsilon$ &$U$ \\
(4) &$\epsilon$ &$U$ &$U'$ \\
 &$U$ &$U'$ &$\epsilon$ \\
\hline
 &$U'$ & &$\epsilon,U$ \\
(5) &$\epsilon$ & &$U',U$ \\
 &$U$ & & $U',\epsilon$\\
\hline
\end{tabular}
\end{center}
\label{tab2}
\end{table}

\end{document}